\documentclass[prl,twocolumn,showpacs,preprintnumbers,amsmath,amssymb,citeautoscript]{revtex4}
\usepackage{graphicx}
\usepackage{dcolumn}
\usepackage{amssymb}
\usepackage{amsmath}
\begin{document}

\title{First-Principles Calculations of Luminescence Spectrum Line Shapes for Defects in Semiconductors: The Example of GaN and ZnO}

\author{Audrius Alkauskas}
\author{John L. Lyons}
\author{Daniel Steiauf}
\author{Chris G. Van de Walle}
\affiliation{Materials Department, University of California, Santa Barbara, California 93106-5050, USA}
\date{\today}

\begin{abstract}
We present a theoretical study of broadening of defect luminescence bands due to vibronic coupling.
Numerical proof is provided for the commonly used assumption that a multi-dimensional vibrational
problem can be mapped onto an effective one-dimensional configuration coordinate diagram. Our
approach is implemented based on density functional theory with a hybrid functional,
resulting in luminescence lineshapes for important defects in GaN and ZnO that show unprecedented
agreement with experiment. We find clear trends concerning effective parameters that characterize
luminescence bands of donor- and acceptor-type defects, thus facilitating their identification.

\end{abstract}

\maketitle

Defects play a key role in the properties of solids.
From the early days of color centers, the study of luminescence and absorption has
been crucial to defect characterization  \cite{Stoneham}. Theoretical efforts to calculate
the broadening of optical transitions at defects due to the interactions with lattice vibrations were
pioneered by Huang and Rhys \cite{Huang_PRCL_1950} and Pekar \cite{Pekar_1950}.
While those theories and their generalizations \cite{Lax_JCP_1952,Stoneham} have been
very successful in describing the shape of experimental optical bands \cite{Stoneham,Reshchikov_JAP_2005},
this inevitably required the use of empirical fitting parameters. Theory has thus been limited in its
ability to aid the microscopic identification of defects or produce accurate predictions.

In this Letter we report that unprecedented precision can now be achieved by rigorously mapping the
multi-dimensional vibrational problem onto an effective one-dimensional configuration coordinate diagram,
combined with advanced electronic structure techniques
\cite{VanDeWalle_JAP_2004,Heyd_JCP_2003}. We demonstrate the power of the approach with the example of
a number of defects in GaN \cite{Reshchikov_JAP_2005} and ZnO \cite{Ozgur_JAP_2005},
two technologically crucial wide-band-gap semiconductors. Excellent agreement with experiment is achieved
for well-characterized defects, and new insights into vibronic coupling emerge.

Our electronic structure calculations are based on density functional theory using
the hybrid functional of Ref.\ \cite{Heyd_JCP_2003} in the \textsc{vasp} code \cite{VASP}.
The fraction $\alpha$ of the screened Fock exchange admixed to the semilocal exchange was set to 0.31
for GaN and 0.36 for ZnO to reproduce experimental band gaps (3.5 eV and 3.4 eV). By describing
bulk electronic structure  better and reducing self-interaction errors,
hybrid functionals substantially improve the accuracy of defect calculations
\cite{Pacchioni_PRB_2000,Alkauskas_PRL_2008,Lyons_APL_2009}. Defects were treated via the supercell
approach \cite{VanDeWalle_JAP_2004}, the interaction with nuclei was described within the
projector-augmented wave formalism \cite{VASP}, and electron wave functions were expanded in plane
waves with a cutoff of 400 eV. Normal modes and frequencies have been calculated using finite
differences.

\begin{figure}
\includegraphics[width=6.8cm]{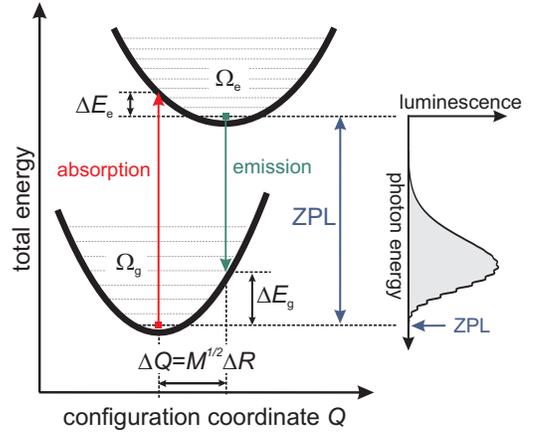}
\caption{(Color online) 1D configuration coordinate diagram describing optical absorption and emission
at a point defect. The minima of the ground-state and excited-state potential energy surfaces are displaced.
$\Delta E_{\{\text{e,g}\}}$ are the relaxation energies and $\Omega_{\{\text{e,g}\}}$ the effective phonon frequencies.
ZPL indicates the zero-phonon line, i.e. the transition between the zero-point vibrational
states in excited and ground-state configurations.
}
\label{1D}
\end{figure}

We illustrate the methodology with the example of the Mg$_{\text{Ga}}$ acceptor in GaN,
a crucial defect since it is the only acceptor impurity capable of making the material
$p$ type. While {\it electrically} acting as a shallow impurity with modest ionization
energy, {\it optically} Mg$_{\text{Ga}}$ behaves as a deep center \cite{Lyons_PRL_2012}:
recombination of an electron at the conduction-band minimum (CBM) with a hole
localized on the neutral Mg$_{\text{Ga}}^0$ acceptor gives rise
to a broad blue luminescence band \cite{Reshchikov_JAP_2005,Lyons_PRL_2012}.
The calculated zero-phonon line (ZPL) energy (see Fig.~\ref{1D}) is 3.24 eV.
We assume here that optical transitions start with a delocalized charge carrier;
excitonic effects are small \cite{Shan_PRB_1996}.

The general theory of luminescence was described in Refs.\ \cite{Stoneham,Huang_PRCL_1950,Lax_JCP_1952}.
When optical transitions are dipole-allowed, as is the case for the defects studied in this work,
at $T$=$0$ the normalized luminescence intensity (lineshape) in the leading order
(the Franck-Condon approximation) can be written as $G(\hbar\omega)=C\omega^3A(\hbar\omega)$,
where $A(\hbar\omega)$ is the normalized spectral function
\begin{equation}
A(\hbar\omega)=\sum_{n} \left|\left<\chi_{\text{e}0}|\chi_{\text{g}n}\right>\right|^2
\delta\left(E_{\text{ZPL}}-\hbar\omega_{\text{g}n}-\hbar\omega\right),
\label{SpectralFun}
\end{equation}
and $C^{-1}=\int A(\hbar\omega) \omega^3 d(\hbar\omega$).
The sum runs over all vibrational levels with energies $\hbar\omega_{\text{g}n}$ of the ground state,
$\chi$ are ionic wavefunctions, and $E_{\text{ZPL}}$ is the energy of the ZPL.
Generalization to finite $T$ is straightforward.

Evaluation of $A(\hbar\omega)$ is complicated by the fact that,
first, the sum includes all relevant vibrational degrees of freedom, and second, normal modes
$\mathbf{Q}_{\text{e}}$ and $\mathbf{Q}_{\text{g}}$ 
in the excited and the ground state are usually 
not identical. The two are related via the Duschinsky transformation $\mathbf{Q}_\text{e}=\mathbf{J}\mathbf{Q}_{\text{g}}+\Delta \mathbf{Q}$
\cite{Duschinsky_AP_1937}, and $\left< \chi_{\text{e}0} | \chi_{\text{g}n} \right>$ are thus highly-multi-dimensional integrals.
For small molecular systems recursive techniques to calculate such integrals have been developed \cite{Doktorov_JMS_1977}
and implemented \cite{Borrelli_JCP_2003}. The large number of vibrational modes that occur for defects in solids
render such a direct approach computationally prohibitive.

Broad optical bands have most often been described via 1D configuration-coordinate diagrams (CCDs)
\cite{Stoneham,Lax_JCP_1952} [Fig.\ \ref{1D}], based on the assumption that the
large number of vibrational modes (with different frequencies) contributing to the lineshape can be
replaced by a single effective mode (sometimes a small number of modes).
The parameters entering the 1D model are the modal mass $M$ of the effective
vibration, the displacement of the potential energy minima $\Delta R$, and the effective frequencies
$\Omega_{\text{g}}$ and $\Omega_{\text{e}}$ [Fig.\ \ref{1D}]. Based on these, the widely used ``Huang-Rhys (HR) factors''
\cite{Huang_PRCL_1950} are defined as the average number of phonons created during a vertical transition:
$S_{\text{g}}=\Delta E_{\text{g}}/\hbar\Omega_{\text{g}}$ and $S_{\text{e}}=\Delta E_{\text{e}}/\hbar\Omega_{\text{g}}$.
There are many examples where a 1D model with empirical fitting parameters provides a good approximation to experimental luminescence lineshapes \cite{Stoneham,Reshchikov_JAP_2005}; still, because it is strictly valid only when all the modes have the same frequency
\cite{Huang_PRCL_1950}, its general applicability has often been questioned.
More importantly, the use of fitting parameters precludes linking to
potentially valuable microscopic information about the defect and limits the predictive power.

Here we address this problem using the following strategy.
Vibrations that couple strongly to the distortion of the geometry
are expected to be dominant in $A(\hbar\omega)$. Such modes have finite
weight on the atoms that experience the largest relaxations, i.e.,
the atoms close to the defect.  When only a small number of atoms
are included, $\left< \chi_{\text{e}0} | \chi_{\text{g}n} \right>$ can be evaluated exactly, 
taking mode mixing into account \cite{Doktorov_JMS_1977,Borrelli_JCP_2003}. 
This exact evaluation can then serve as a test of the accuracy of any approximations.
Once an approximate treatment has been validated in this fashion, it can be applied
to much larger systems of atoms provided it is sufficiently less numerically demanding
than the exact evaluation.

\begin{figure}
\includegraphics[width=8.5cm]{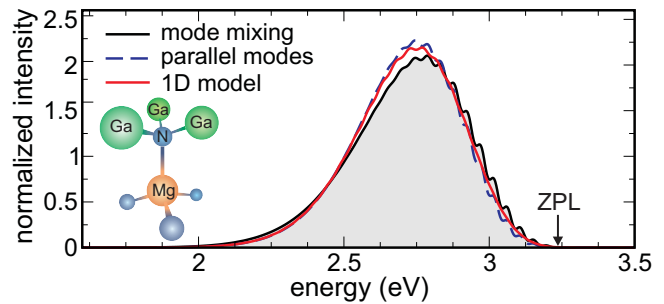}
\caption{(Color online) Normalized luminescence spectrum of the Mg$_{\text{Ga}}$(0/$-$) transition,
calculated taking into account vibrational modes of only those atoms that relax most 
(labelled in the inset).
Black solid line (and shaded area): calculation including mode mixing; blue dashed line: parallel-mode approximation;
red solid line: effective 1D vibrational problem.
}
\label{Mg1Dq-1}
\end{figure}

In the case of Mg$_{\text{Ga}}^0$ a hole is localized on a N neighbor of the Mg atom, and
the five atoms surrounding this hole account for more than 90\% of the whole relaxation.
The calculated luminescence lineshape, taking mixing between the resulting 15 vibrations into account,
is shown in Fig.\ \ref{Mg1Dq-1}. Multi-dimensional overlap integrals were calculated using the \textsc{molfc} code
\cite{Borrelli_JCP_2003}. We have applied a Gaussian smearing with a small $\sigma$=$0.01$ eV to simulate
additional broadening mechanisms, resulting in a smooth lineshape.

Recursive algorithms \cite{Doktorov_JMS_1977} lead to exploding computational requirements when applied
to larger atom clusters. A simplification that is often used for molecules \cite{Borrelli_JCP_2003} and
almost always implied in solids \cite{Lax_JCP_1952} is the parallel-mode approximation, in which the
eigenmodes of either the ground state or the excited state are chosen as common vibrational states,
leading to  $\mathbf{J}$=$\mathbf{1}$. $\left< \chi_{\text{e}0} | \chi_{\text{g}n} \right>$
then factorizes into 1D integrals, each corresponding to one vibrational mode, greatly reducing computational complexity.
As seen in Fig.\ \ref{Mg1Dq-1}, the resulting lineshape is indeed close to the exact result. Therefore,
while mode mixing is present, it is not substantial. 

Now that we have validated the parallel-mode approximation we can include more atoms, since overlap
integrals become easy to calculate. However, the number of terms that have to be included
grows very rapidly with system size.
This reflects the fact that important modes often do not occur in the gap of the bulk
phonon spectrum but are resonances, and therefore not well localized in
real space. Consequently, further approximations are required. This we achieve by devising
a suitable 1D CCD based on computed parameters as outlined below.

The weight by which each normal mode $k$ contributes to the distortion of the defect geometry
during optical transition can be written as
$p_k=(\Delta Q_{k}/\Delta Q)^2$, where
\begin{equation}
\Delta Q_{k} = \sum_{\alpha i} m_{\alpha}^{1/2} \Delta R_{\alpha i} q_{k; \alpha i};
\left(\Delta Q\right)^2 = \sum_{k} \Delta Q_{k}^2 \, .
\label{expansion-modes}
\end{equation}
Here $\alpha$ labels atoms, $i$=$\{x,y,z\}$, $\Delta R_{\alpha,i}=R_{\text{e};\alpha,i}-R_{\text{g};\alpha,i}$
is the distortion vector, $R_{\{e,g\};\alpha,i}$ are atomic coordinates, and $q_{k; \alpha i}$ is the
unit vector in the direction of the normal mode $k$ 
$\left(\sum_{\alpha i} q_{k; \alpha i}q_{l; \alpha i} = \delta_{k,l}\right)$.
We find that it is useful to define an effective frequency
\begin{equation}
\Omega^2_{\{\text{e},\text{g}\}} = \left< \omega_{\{\text{e},\text{g}\}}^2 \right> = \sum_{k} p_{\{\text{e},\text{g}\};k} \omega_{\{\text{e},\text{g}\};k}^2,
\label{Omega-1}
\end{equation}
where $\omega_{\{\text{e,g}\};k}$ is the frequency of the mode $q_{\{\text{e,g}\};k}$.

Parameters $E_{\text{ZPL}}$, $\Delta Q$, $\Omega_{\text{g}}$ and $\Omega_{\text{e}}$ define
a 1D CCD (cf.\ Fig.\ \ref{1D}) for a quantum oscillator with unit mass and can be used to
calculate the luminescence lineshape. 
Gaussian smearing is still applied, but it should now reflect the replacement 
of many vibrations at various frequencies with one effective frequency.
Inspection of the mean-square deviation for the distribution of phonon frequencies that contribute to the distortion
leads to $\sigma$=$0.025$ eV ($\sim0.6\hbar\Omega_{\text{g}}$). 
The result for Mg$_\text{Ga}$
in Fig.\ \ref{Mg1Dq-1} shows that our rigorously defined 1D model is an excellent approximation
to the multi-dimensional calculations.

Now that the validity of the 1D model has been established,
it can be extended to larger numbers of atoms and we no longer need to explicitly
calculate normal modes and frequencies to determine the effective parameters $\Delta Q$
[Eq.\ (\ref{expansion-modes})] and $\Omega_{\{\text{e},\text{g}\}}$ [Eq.\ (\ref{Omega-1})].
Indeed, by inserting the expression for $\Delta Q_k$ into the one for $\Delta Q$
in Eq.\ (\ref{expansion-modes}), one can show that when all the atoms in the supercell
are included in the vibrational problem, $\left(\Delta Q\right)^2 = \sum_{\alpha, i} m_{\alpha} \Delta R_{\alpha i}^2$.
The modal mass is defined via $\Delta Q = M^{1/2}\Delta R$, where
$\left(\Delta R\right)^2 = \sum_{\alpha, i}\Delta R_{\alpha i}^2$.
Effective frequencies $\Omega$ can be obtained by mapping the potential energy surface
around the respective equilibrium geometries along the path that linearly interpolates between the two geometries \cite{Schanovsky_JVST_2011}.
A third-order polynomial fit was found to suffice in all cases.
The frequency $\Omega$ in the quadratic term defined in this way is equivalent to the
one calculated from Eq.\ (\ref{Omega-1}). Third-order anharmonic corrections 
that affect vibrational wavefuntions and thus overlap integrals,
were included in the calculations of the spectral function perturbatively. 
For all subsequent calculations, we have used
96-atom wurtzite supercells, and relaxations of all the atoms were included
in determining effective parameters. 
The resulting parameters are summarized in Table \ref{tableI}.

\begin{table*}
\caption{Effective parameters for various defect-related luminescence transitions in GaN and ZnO.
$\Delta Q$ and $\Delta R$: total mass-weighted and total distortions;
$M$: modal mass; $\Omega_{\{\text{e,g}\}}$: effective frequencies in the ground
and excited states (charge state in parentheses); $
E_{\text{ZPL}}$: zero-phonon line energy; 
FWHM: full-width at half maximum
of the band;
T: temperature for which the FWHM is given; 
$S_{\text{\{e,g\}}}$: Huang-Rhys factors.
If experimental parameters are not explicitly given in the corresponding experimental papers, they are extracted using the original
data and are shown in italics.
}
\begin{ruledtabular}
\begin{tabular}{l c c c c c c c c c c c}
Defect, charge states,                    &         &$\Delta Q$      & M    & $\Delta$R &$\hbar\Omega_{\text{g}}$&$\hbar\Omega_{\text{e}}$&$E_\text{ZPL}$&FWHM at T&$S_\text{g}$&$S_\text{e}$&\\
and optical transition                    & Method  &(amu$^{1/2}$\AA)& (amu)& (\AA)&(meV)              &(meV)& (eV)        &(eV, K)&&&\\
\hline
Mg$_{\text{Ga}}$ (0/$-$) in GaN& theory  &1.6       &45  &0.24 &47 ($-$) &34 (0)     &3.24 \cite{Lyons_PRL_2012}  &0.44 at 0&12       &-&\\
Mg$_{\text{Ga}}^0+\text{e}^{-}\rightarrow$ Mg$_{\text{Ga}}^-$& expt. &  &      &     &  &   &\emph{3.30} \cite{Reshchikov_PRB_1999}&\emph{0.36} at 13 \cite{Reshchikov_PRB_1999}&       &\vspace{1mm}&\\
\hline
C$_{\text{N}}$ (0/$-$) in GaN            & theory  &1.6            &51  &0.22 &42 ($-$)&36 (0)   &2.60 \cite{Lyons_APL_2010}  &0.35 at 77 &11       &10&\\
C$_{\text{N}}^0+\text{e}^{-}\rightarrow$C$_{\text{N}}^{-}$& expt.   &                &      &     &41$\pm$5    &40$\pm$5 \cite{Ogino_JJAP_1980}     &2.64 \cite{Ogino_JJAP_1980}  &\emph{0.39} at 77 \cite{Seager_JL_2004}&12.8$\pm$1.6       &13.4$\pm$1.7 \cite{Ogino_JJAP_1980}\vspace{1mm}&\\
\hline
$V_{\text{N}}$ (+3/+2) in GaN            & theory  &3.7            &68  &0.45 &23 (+2) & 21 (+3)&3.02 \cite{Yan_APL_2012} &0.33 at 0&  36     &35&\\
$V_{\text{N}}^{+3}+\text{e}^{-}\rightarrow V_{\text{N}}^{+2}$&expt.  &                &      &     &          &         &\emph{3.07} \cite{Salviati_MRS_2000}  &\emph{0.36} at 5 \cite{Salviati_MRS_2000}&    &&\\
\hline
N$_{\text{O}}$ (0/$-$) in ZnO            & theory  &1.9            &48  &0.28 &40 ($-$) & 32 (0)&2.20 \cite{Lyons_APL_2009}& 0.54 at 300 &15&15&\\
N$_{\text{O}}^{0}+\text{e}^{-}\rightarrow$ N$_{\text{O}}^{-1}$&expt.  &                &      &     &          & &\emph{2.30} \cite{Tarun_AIP_2011}&\emph{0.55} at 300 \cite{Tarun_AIP_2011}&& &\\
\end{tabular}
\label{tableI}
\end{ruledtabular}
\end{table*}

\begin{figure}
\includegraphics[width=8.5cm]{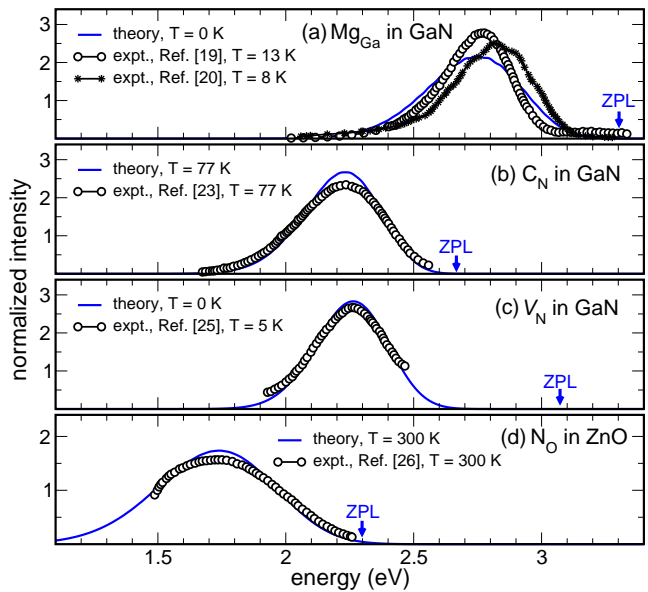}
\caption{(Color online) Calculated (solid lines) and measured (symbols) luminescence lineshapes for various defects in GaN and ZnO.
(a) Mg$_{\text{Ga}}$ in GaN, expt.\ from Refs.\ \cite{Reshchikov_PRB_1999} (disks) and \cite{Leroux_JAP_1999} (stars)
(b) C$_{\text{N}}$ in GaN, expt.\ from Ref.\ \cite{Seager_JL_2004};
(c) $V_{\text{N}}$ in GaN, expt.\ from Refs\ \cite{Salviati_MRS_2000}.
(d) N$_{\text{O}}$ in ZnO, expt.\ from Ref.\ \cite{Tarun_AIP_2011}
The arrows indicate the ZPL, and the calculated spectra were shifted by $\sim$0.1 eV, as discussed in the text.}
\label{spectra}
\end{figure}

The luminescence lineshape $G(\hbar\omega)$ for Mg$_{\text{Ga}}$ at $T$=$0$ K
is shown in Fig.\ \ref{spectra}(a), together with experimental low-temperature data from Refs.\
\cite{Reshchikov_PRB_1999} and \cite{Leroux_JAP_1999}. To facilitate comparison with experiment,
the calculated lineshapes were shifted to bring the maximum of the luminescence in agreement with that of the
experimentally measured curves. The magnitude of the shift provides an estimate for the error in the $E_\text{{ZPL}}$ 
(thermodynamic transition level), and is less than 0.1 eV for all defects. This error bar reflects both the 
remaining inaccuracy of even the most advanced first-principles methods, and any electrostatic corrections 
(such as excitonic effects or donor-acceptor interactions) not included in the present model.  
The width of the theoretical band (0.44 eV) is only slightly larger
than the experimental value of 0.36-0.37 eV. We derive the effective vibrational frequency in the ground state
$\Omega_{\text{g}}$=47 meV, and the HR factor $S_{\text{g}}$=12. The good agreement between the calculated
and experimental lineshapes for Mg$_{\text{Ga}}$ attests to the power of the approach presented here, when used
in combination with state-of-the-art density functional calculations.

The agreement with experiment is even better for our second example, C$_{\text{N}}$.
This defect has been suggested as a source of yellow luminescence (YL)
\cite{Ogino_JJAP_1980,Seager_JL_2004,Reshchikov_JAP_2005} based on the transition
C$_{\text{N}}^0$+e$^-$$\rightarrow$C$_{\text{N}}^{-}$ \cite{Lyons_APL_2010} (where e$^-$ is an electron at the CBM).
This defect again exhibits hole localization in the neutral charge state, but now
the hole is localized on the C atom \cite{Lyons_APL_2010}. The calculated luminescence lineshape at $T$=$77$ K
[Fig.\ \ref{spectra}(b)] agrees very well with the measurements of Ref.\ \cite{Seager_JL_2004},
in which YL was convincingly attributed to a C-related defect. Our calculated effective frequencies
($\hbar\Omega_{\text{\{e,g\}}}$=36, 42 meV) and HR factors ($S_{\text{\{e,g\}}}$=10, 11) are in an
excellent agreement with those
measured in Ref.\ \cite{Ogino_JJAP_1980} (Table \ref{tableI}). 

Our next example is $V_{\text{N}}$. Nitrogen vacancies have low formation energies in $p$-type GaN,
and they have been suggested \cite{Yan_APL_2012} as a cause of the YL in $p$-GaN observed in Ref.\ \cite{Salviati_MRS_2000}.
The calculated luminescence lineshape for the transition $V_{\text{N}}^{+3}$+e$^-$$\rightarrow V_{\text{N}}^{+2}$
[Fig.\ \ref{spectra}(c)] shows impressive agreement with these low-temperature experiments.

To demonstrate that the approach is not limited to GaN, in Fig.\ \ref{spectra}(d) we compare the calculated and
measured \cite{Tarun_AIP_2011} luminescence lineshapes at $T$=$300$ K for the deep N$_{\text{O}}$ acceptor
in ZnO, corresponding  to the transition N$_{\text{O}}^{0}$+e$^-\rightarrow$N$_{\text{O}}^{-1}$
\cite{Lyons_APL_2009}.  The agreement between theory and experiment is again extremely good. 

Our ability to calculate accurate parameters allows us to examine some general trends. 
For the two substitutional acceptors in GaN analyzed above,
total distortions $\Delta R$ amount to $0.22-0.24$ \AA, and HR factors to $10-12$,
irrespective of whether the hole is bound to C or N.
Other acceptors with anion-bound holes ($V_{\text{Ga}}$, Be$_{\text{Ga}}$, Zn$_{\text{Ga}}$) show very similar 
behavior. The donor $V_{\text{N}}$, on the other hand, is very different. Its defect 
wave function is composed mainly of Ga 4$s$ states, and $\Delta R$ associated with the (+3/+2) transition is 0.45 \AA,
almost twice as large as in the case of acceptors.  The distortion mostly affects the four nearest Ga atoms,
leading to a large modal mass and small effective frequencies (Table \ref{tableI}). 
In conjunction with large relaxation energies (0.72 and 0.82 eV) this results in very large HR factors ($S_{\text{\{e,g\}}}=35,36$).
This is in contrast with the acceptors, where anions are involved in the distortion, leading to a smaller modal masses, higher 
effective vibrational frequencies, and hence smaller HR factors. Similar trends are observed for ZnO:
we find $\hbar\Omega_{\{\text{e,g}\}}$=28-40 meV and $S_{\{\text{e,g}\}}$=15-23 for acceptors with anion-localized holes
(N$_\text{O}$, $V_\text{Zn}$, Li$_\text{Zn}$) while $\hbar\Omega_{\{\text{e,g}\}}$=16, 21 meV and $S_{\{\text{e,g}\}}$$\approx$50 for 
donors with cation-derived states (V$_{\text{O}}$). The general result for acceptors in ZnO is in
accord with experimental data of Ref.\ \onlinecite{Reshchikov_2006}.

While such {\it a posteriori} interpretations are simple and intuitive, they are only reliable if based on an accurate microscopic 
description of the defect.  We note that model calculations have yielded results that were very different from our first-principles 
values (e.g., $S$=3.5$-$6.5 for defects related to YL in Ref.\ \cite{Ridley_JPCM_1998}), 
starkly illustrating the shortcomings of such approaches.

High values of HR factors mean that it is very difficult to determine the ZPL
in experimental luminescence spectra. Indeed, the weight of the ZPL is expotentially 
supressed for larger $S$: $\left|\left<\chi_{\text{e}0}|\chi_{\text{g}0}\right>\right|^2\approx\exp\{-S_{\{\text{e,g}\}}\}$
(equality holds for $\Omega_{\text{g}}=\Omega_{\text{e}}$ \cite{Huang_PRCL_1950}).
As seen in Fig.\ \ref{spectra}, such complication arises for all the defects studied 
here and is especially aparent for $V_{\text{N}}$. This highlights the practical use of
calculations exemplified in the current work.

The examples have demonstrated that our methodology is capable of producing luminescence
lineshapes in very good agreement with experiment, as well as 
quantities  that can be directly compared with experimental parameters.
The achieved agreement is based, of course, on the accuracy of the underlying electronic
structure method, but also on the applicability of the 1D model to broad luminescence bands.
While the explicit consideration of many vibrational
modes is sometimes needed to understand the
experimental spectra when $S\approx01$ (i.e., defects with moderate electron-phonon coupling) 
\cite{Stoneham,Lax_JCP_1952,Kretov_JL_2012},
We have demonstrated that such a model, with suitably calculated parameters, is indeed valid
for defects with large electron-phonon coupling ($S\gg1$),
even if many phonon modes couple to the optical transition. Another important conclusion
that follows from our work is that the effective mode frequency is usually
much smaller than that of LO phonons (91 meV in GaN and 73 meV in ZnO), contrary to what has been
commonly assumed in phenomenological approaches \cite{Huang_PRCL_1950}. 
Our conclusion is in full agreement with detailed experimental measurements for color
centers in alkali halides \cite{Russel_PR_1956}, indicating the generality of
the obtained result.

Our findings are important for future studies of semiconductors and insulators that exhibit broad defect luminescence bands.
In particular, as our examples showed the methodology provides a means for identifying the microscopic origin of the 
numerous as yet unassigned luminescence bands
in technologically important wide band-gap materials. More generally, our work attests to the success 
of first-principles methods to describe electron-phonon interactions in solids \cite{Giustino_PRB_2007}
beyond the application to perfect crystals.

We thank F. Bechstedt, N. Grandjean, A. Janotti, M. A. Reshchikov,
and Q. Yan for discussions, as well as A. Borelli for a copy of the
\textsc{molfc} code.
AA acknowledges a grant from the Swiss NSF (PA00P2$\_$134127).
Additional support was provided by NSF (DMR-0906805), and
computational resources  by XSEDE (NSF DMR07-0072N) and
NERSC (DOE Office of Science DE-AC02-05CH11231).


\end{document}